# [NiFe] dithiolene diphosphine complex for hydrogen gas activation: a Theoretic Insight.

Jing Gu-Yan*, Steve Y. Bollinger

Department of Chemistry, Princeton University, Princeton, NJ 08540

*Supporting Information Placeholder*

**ABSTRACT:** A diphosphino-nickel-iron dithiolene complex, [Ni(bdt)(dppf)] (bdt = 1,2-benzenedithiolate, dppf = 1,1-bis(diphenylphosphino)ferrocene), has been recently found to be reasonably active on proton reduction to dihydrogen (*J. Am. Chem. Soc.* **2015**, *137*, 1109). Interestingly, this exceptional complex was found to be also reactive towards dihydrogen activation as indicated by the electrochemical investigation. However, a pure nickel dithiolene diphosphine theoretical mode, excluding the contributions from iron moiety, was applied to attribute the experimental catalytic observation. We have re-visited the theoretical approach in details for this [NiFe] catalyst and compared it with the non-active nickel dithiolene diphosphine complexes. We found that both nickel and iron moieties in this newly developed complex were imperative for the observed catalytic performance, particularly towards the activation of dihydrogen.

Renewable energy approaches, including catalytical generation of hydrogen fuel from water;[1] production of formate, alcohols, hydrocarbons or other energy-dense carbon-based compounds from carbon dioxide reductions in aqueous solutions,[2] have recently spurred scientists' interest. One of major challenges to overcome the applicable obstacle of these approaches is to develop an inexpensive but also effective catalyst. Interestingly, nature did reward us with such system, for example, hydrogenases, which only contain the first row transition metals iron and nickel, can operate under very lower overpotential and reduce proton to hydrogen reversibly in weakly acidic aqueous solutions at very high turnover frequencies (TOF), more than 1000 s$^{-1}$.[3,4] Chemical mimic syntheses of nickel-iron complexes have been extremely popular for the past decade, particularly after the elucidation of the [NiFe] hydrogenase enzyme structure by Volbeda et al. in 1995.[5]

Very recently, Gan *et al.* have successfully developed an extremely effective NiFe catalyst,[6] compound **1**, [Ni(bdt)(dppf)] (bdt = 1,2-benzenedithiolate, dppf = 1,1-bis(diphenylphosphino)ferrocene) (Figure 1), for proton reduction with TOF 1240 s$^{-1}$, even higher than the authentic biological enzymatic NiFe biocatalyst, noted as 1000 s$^{-1}$.[3,4] It is imperative to point out that this complex also exhibited hydrogen gas activation in which the quasi-reversible redox waves of this complex were observed to shift anodically for ca. 270 mV if hydrogen gas was present in the solution. Simultaneously, a noticeable decrease in the reversibility of the process was observed in the presence of hydrogen where a diminished re-oxidation wave was observed as resulted from the ratio of oxidation to reduction current to only ca. 70% to that case of the absence of hydrogen. Interestingly, the complex **2**, [Ni(bdt)(dppe)] (dppe=1,2-bis(diphenylphosphino)ethane) was reported to be inactive for any catalytic observation, either proton reduction or dihydrogen activation. However, these authors believe that "the reactivity of **1** is entirely dependent on the Ni atom. Thus, **1** is more analogous to mononuclear Ni or Co catalysts than to other [NiFe] complexes with a metal−metal bond." We have theoretically investigated the related complexes and found that iron moiety is also of critical importance, particularly towards the hydrogen gas activation.

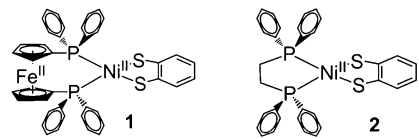

**Figure 1.** Chemical structures of complex **1** and **2**.

The structural and bonding properties metallo-dithiolene complexes[8] and organic dithiolate complexes,[7] have been previously extensively studied. However, such nickel diphosphine dithiolene complex's theoretical investigation directly targeting proton reduction has been of fewer studies. To our knowledge, there has been no report of dihydrogen gas activation or oxidation with dithiolene complexes. Therefore, this work focused on nickel-iron complex's electrochemical activation reaction with dihydrogen gas.

It has been reported that a significant shift (270mV, anodic) in the redox waves of complex **1** in cyclic voltammetry was occurred when hydrogen gas was introduced into the electrolyte, indicating an interaction between complex **1** and hydrogen.[6] The reversibility of the redox wave is quite diminished, thus implying activation of dihydrogen might be possible. However, spectroscopic evidence of neither **1-H$_2$** nor **(1-H$_2$)$^-$** can be observed. Moreover, hydrogen oxidation products, even with the presence of base that would be able to shift the reaction equilibrium towards products, have not been observed. Therefore, it has been concluded the dihydrogen gas is involved in the electrochemical reaction with complex **1**, however, this interaction or activation has not yet oxidative enough, at least in the chemical pathway, to produce proton from dihyrogen under these conditions.

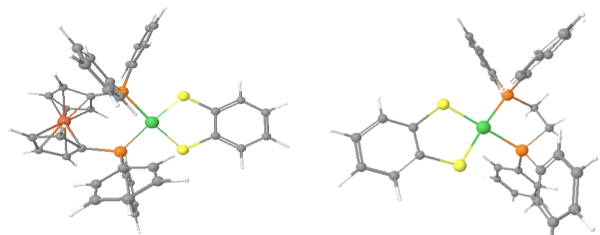

**Figure 2.** Optimized structures of **1** (left); **2** (right). Atom color: green-nickel; red-iron; yellow-sulfur; orange-phosphine; gray-carbon; white-hydrogen.

The theoretical approach was achieved based on density functional theory (DFT) according to previous published method as detailed[8b,c,9] in following: Geometry optimizations of these complexes were performed with the Gaussian 09 suite of software and employed the B3LYP functional. The authenticity of each converged structure was confirmed by the absence of imaginary vibrational frequencies. A double-ζ (DZ) basis set with an effective electron core potential (LANL2DZ ECP) was used for nickel and iron, a triple-ζ basis with two polarization functions was used for sulfur and phosphine, and basis sets 6-31(d,p) was applied for the remaining atoms. The polarizable continuum model (PCM) was applied to model solvent effects.

**1** + e$^-$ → **1**$^-$      $\Delta$G, E$_{1/2}$     (1)

**1 – H$_2$** + e$^-$ → **(1-H$_2$)$^-$**      $\Delta$G, E$_{1/2}$     (2)

$$E_{1/2} = -\frac{\Delta G^o_{1/2}}{n_e F} - E^{o,ref}_{1/2} \quad (3)$$

The possible reduction reactions were illustrated in equation 1-2. Redox potential computations of equation 3 require at least one empirical value. An absolute value for the reference electrode, for instance, SCE (saturated calomel electrode) as indicated in the reference 6 to illustrate the electrochemistry of complex **1** and **2**, is needed in order to compute the energy for the electron in solution. It is widely accepted that corrections between different reference electrodes and the SHE (standard hydrogen electrode) are known with very high accuracy. However, the absolute value for the SHE always presents a significant variance in empirical values.[10] If such empirical value is required, herein we will utilize the value given by Isse etc.,[10] noted as 4.281 V, because it uses the IUPAC derivation with revised energy components.

Our theoretical data on optimized structure of **1** and **2** is corroborated well with the experimental crystal structures as indicated in Figure 2 and Table 1. Approximately 0.067Å longer of Ni-P bond in **1** was observed than in **2**, this value is 0.065 from experimental bond length comparison (reference 11 for crystal structure of **2**). The Ni−S bond lengths are computed about 0.023 Å longer for **1** than **2** as well, while crystal structurally, it has ca. 0.022 Å difference. In general, the nickel-iron complex exhibits a relaxed structure with lower bond order, providing more metallic valence electrons available for extra bonding possibility, i.e. hydrogen, than complex **2**. The Ni-Fe distance was calculated about 0.042 Å longer than crystal structure, within the error of theoretical predications between solid-state crystal structure and solvated complexes. Hence, the theoretical methods in this work are reliable to predicate the structure information of these complexes and thus they will be further applied for electrochemical investigation.

**Table 1.** Experimental (crystal structures, reference 6 and 11) atomic distances and calculated theoretical (THF solvation) atomic distances (Å).

| Distance | 1 Exp. | 1 Theo. | 2 Exp. | 2. Theo. |
|---|---|---|---|---|
| Ni-P | 2.225 | 2.239 | 2.160 | 2.172 |
| Ni-S | 2.166 | 2.172 | 2.144 | 2.149 |
| Ni-Fe | 4.257 | 4.299 | - | - |

The one electron reduced complexes **1$^-$** and **2$^-$** have also been computed. Gibbs free energy was calculated at standard condition, directly obtaining from the vibrational frequencies computations. With the empirical value for SHE, the redox potentials were thus predicated from Equation 3. In order to validate this method, we have also conducted a calculation using ferrocene (with the same theoretical method) as an internal comparison molecule. The obtained redox potentials are listed in Table 2 and compared with the experimental data. Surprisingly, for ferrocene, the calculated redox potential is only 90 mV lower than the experimental observation[12] in THF when tetrabutylammonium hexafluorophosphate was applied as the electrolyte. The calculated redox potentials for complex **1** and **2** reasonably corroborated with the experimental data with only ca. 0.1V error. This method provides a trustworthy pathway to predicate the redox poten-tial of these complexes. Interestingly, the calculated redox potentials from the list, are all ca. 0.1 V lower than the experimental value, implying a highly constant trend of these calculations. More importantly, if we were able to apply the ferrocene calculation as the internal correction, the obtained corrected redox potential for complex **1** and **2** as listed in Table 2 are matching excellently with the experimental data within only marginal difference. The error of redox potential calculations for complex **2** in THF under current method is within error of 4%. Probably due to the similar structure of **1** and ferrocene, extraordinarily, the error of complex is even less than 1%.

**Table 2. Redox potentials (vs. SCE) obtained from experimental and theoretical approach.**

| Complex | Experimental observation* | Theoretical calculation | Corrected potential** |
|---|---|---|---|
| **1** | -1.280 | -1.38 | -1.29 |
| **2** | -0.518 | -0.63 | -0.54 |
| **Ferrocene** | 0.56 | 0.47 | - |

* From reference 6 and 12. ** Corrected value was obtained by adding -0.09 V for each theoretical value. (-0.09V = 0.47V-0.56V)

With these highly reliable structural calculations and redox potential predications, we are targeting to investigate the dihydrogen's reaction with these complexes. Different from proton's bonding with possible reactive sites on either sulfur or nickel atoms, dihydrogen binding can be only achieved through metallic center. For complex **2**, no structure of dihydrogen-bonded complexes can be obtained after exclusive exploration from the initial structures when hydrogen molecules are placed near nickel center. Instead, complex **2** and dihydrogen molecule are optimized separately, implying zero stabilization energy between these two. It is noteworthy to point out, for the reduced complex **2⁻**, a similar result with **2** was obtained. In other words, no binding interaction between complex **2** or **2⁻** with dihydrogen can be obtained. This is consistent with the experimental observation that no electrochemical properties change can be observed in the presence or absence of hydrogen gas in solution for complex **2**.

However, the binding reactions for complex **1** are more complicated. If the iron center was completely ignored as Gan *et al.* applied for proton reductions, in other words, the dihydrogen molecule was initially positioned above or below the square plan of $NiS_2P_2$. To our surprise, there has been no obvious difference of such binding natures as it to complex **2**, although a more relaxed structure with longer Ni-P and Ni-S bonds was obtained for **1**. Under such binding strategy, zero binding energy was obtained for complex **1** as well. The electronic influence between dppe and dppf ligand to the nickel center, which might result in the redox potential [assigned as Ni(II)/Ni(I)] difference of ca. 0.8V both on both experimental and theoretical level, however, such electronic effect from the ligand on nickel center is not responsible for the dihydrogen binding from our DFT calculations.

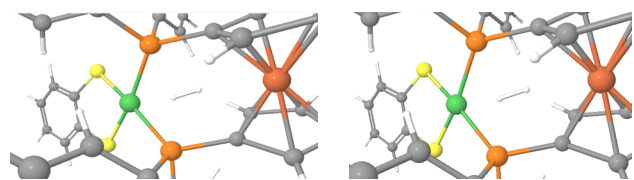

**Figure 2.** Optimized structures of **1-H₂** (left); **(1-H₂)⁻** (right). Hydrogen atoms distance in the left: 0.749Å, right: 0.823Å. Atom color: green-nickel; red-iron; yellow-sulfur; orange-phosphine; gray-carbon; white-hydrogen.

Therefore it is highly motivated for us to further explore the dihydrogen binding between the nickel and iron center. An indeed optimized structure for complex **1-H₂** was obtained when a hydrogen molecule was initially positioned in line with Ni-Fe atoms. The optimized structure of this binding complex was shown in Figure 2. The dihydrogen binding with one electron reduced complex **1⁻** was also optimized in the similar approach. The binding energy was calculated from the equation 5 and 7 and listed in Table 3. In addition, the Gibbs free energy was also computed from frequency calculations and listed in Table 3. The hydrogen atom distance in **1-H₂** was recorded as 0.750 Å, comparing with the free hydrogen distance 0.744 Å in dihydrogen molecule. Such marginal distance variation is indicating a very small binding energy between **1** and dihydrogen molecule. This is consistent with the binding enthalpy of only -2.9 Kcal/mol for **1-H₂**. Moreover, even a positive Gibbs free energy for the binding reaction was observed, rendering a non-spontaneous reaction between **1** and dihydrogen molecule under the standard condition. Nevertheless, the iron center is critical for the dihydrogen binding, otherwise, the complex **1** and **2**, if nickel moieties are only concerned, resulting in zero binding energy to dihydrogen molecule, and are of no difference between themselves towards hydrogen binding.

$$1 + H_2 \rightarrow 1\text{-}H_2 \quad (4)$$

$$\Delta H(1-H_2) = H(1-H_2) - H(1) - H(H_2) \quad (5)$$

$$1^- + H_2 \rightarrow (1\text{-}H_2)^- \quad (6)$$

$$\Delta H[(1-H_2)^-] = H[(1-H_2)^-] - H(1^-) - H(H_2) \quad (7)$$

**Table 3. Binding Energy for complex 1 and 1⁻ as dihydrogen bridged between Nickel and Iron atoms.**

| Complex | ΔH Kcal/mol | ΔG Kcal/mol * | ΔH≠ Kcal/mol** |
|---|---|---|---|
| **1-H₂** | -1.9 | 4.9 | 29.8 |
| **(1-H₂)⁻** | -9.2 | -2.2 | 17.3 |

* Gibbs free energy, **G** was obtained directly from frequency calculations, while Δ**G** was formulated similar to equation 5 and 7. ** Transition state energy, explored from free hydrogen and free complex to **1-H₂** or **(1-H₂)⁻**

The reduced complex **(1-H$_2$)$^-$** exhibited more promising binding characters. First, the elongation of the hydrogen distance is significant. 0.823 Å was observed for hydrogen distance in **(1-H$_2$)$^-$**, corresponding to 11% longer than it is in free hydrogen molecule. The details of binding structure and parameters focusing on Ni(H$_2$)Fe was shown in supporting information. More interestingly, both the enthalpy and the Gibbs free energy are negative for this binding reaction, indicating a spontaneous process for this reaction. Therefore, from the thermodynamic prospect, we can safely draw a predication that the isolated reduced complex **1$^-$** can react with hydrogen gas under standard condition and the spectroscopic evidence for dihydrogen molecule binding or even the isolation of complex of **(1-H$_2$)$^-$** might be of great possibility. A simple transition state exploration was conducted with QST2 starting from free hydrogen gas and free complex **1** or **1$^-$** to form **1-H$_2$** or **(1-H$_2$)$^-$**. A significant lower transition state energy, up to 12.5 Kcal/mol, for the binding process of **(1-H$_2$)$^-$** was also observed than for **1-H$_2$**. Such low transition state energy for the formation of **(1-H$_2$)$^-$**, noted as 17.3 Kcal/mol, demonstrated a possible lower activation barrier for dihydrogen binding.

At the end, it is important to re-examine the theoretical conclusions with electrochemical behaviors of the dihydrogen binding reaction. Since we have obtained a highly reliable redox potential calculation which lead to less than 1% error of complex **1**. As indicated in scheme 1, the free energy difference of the redox reaction of **1-H$_2$** can be illustrated from the complex **1** after compensating with the binding free energy.

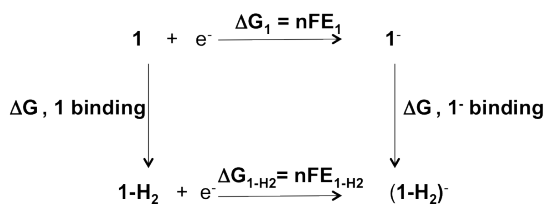

**Scheme 1.** The redox process of **1-H$_2$** derived from complex **1**.

From Table 3, we can conveniently obtain a theoretically computed free energy difference term of -7.1 Kcal/mol between **1-H$_2$** and **(1-H$_2$)$^-$**. This free energy term corresponds to a -0.26V E½ change or anodic shift of the half wave potential, as demonstrated by equation 3, hence from our theoretical approach, the calculated redox potential for the process of reduction **1-H$_2$** to form **(1-H$_2$)$^-$** is ca. -1.03 V vs. SCE. This value is matching surprisingly well with the experimental observed value of -1.009 V by Gan *et al.* as indicated in Table 4. This perfect energy match also indicated our hypothesis of dihydrogen binding in between of Ni-Fe centers is absolutely correct. Therefore, our method provided a meaningful insight and reliable approach to investigate and accommodate the electrochemical behavior of complex **1** with the presence of hydrogen gas.

**Table 4. Redox potentials (vs. SCE) obtained from experimental (reference 6 ) and theoretical approach.**

| Complex | Experimental observation* | Theoretical calculation* |
|---|---|---|
| **1** | -1.280 | -1.29 |
| **1-H$_2$** | -1.009 | -1.03 |

\* After internal correction with ferrocene.

In conclusion, we have theoretically investigated the nickel-iron diphosphine dithiolene complexes' reaction with dihydrogen gas. Our method has successfully generated the structures that are bearing the same key parameters of the complexes' crystal structures. Moreover, we have calculated the redox potential of these complexes, even within 1% of error, after selection of proper empirical value for SHE and ferrocene as internal correction. The dihydrogen binding to complex **2** or its reduced form **2$^-$**, with focus on NiS$_2$P$_2$ square plan, was not observed via our DFT method as demonstrated by zero stabilization energy. Without iron center's binding activity, the complex **1** behaved almost identically to complex **2**, neither **1** nor **1$^-$** was found to bind to dihydrogen. We have proved that the iron's role is as equal importance as nickel for dihydrogen binding. Particularly, a binding energy of 9.2 Kcal/mol of **(1-H$_2$)$^-$** was observed. The free energy for this reduced species is also found to be negative with reasonably low transition state energy; hence we predicated the direct reaction of isolated **1$^-$** with dihydrogen is possible. Finally, we have theoretically corroborated the redox potential from **1-H$_2$** to **(1-H$_2$)$^-$** as evidenced by this method as -1.03 V vs. SCE, which exhibits a prefect match for the electrochemical observation of -1.009 V vs. SCE. Such anodic shift of 0.26 V from our DFT calculation is perfectly observed in the experimental change of 0.27V, indicating our binding approach of dihydrogen between Ni-Fe is true. The binding reactions of dihydrogen gas with these nickel-iron dithiolene diphosphine complexes have thus been well characterized by our DFT methods. This method might be of further significant to explore such multi-metal complexes towards renewable energy investigations.

## ASSOCIATED CONTENT

### Supporting information
Optimized structures both in gas phase and in THF solvent, details of computational methods, QST2 for transition state energy, details of binding structure and parameters focusing on Ni(H$_2$)Fe etc.

## AUTHOR INFORMATION

### Corresponding Author
jingguchem@gmail.com (JG)

**Present Addresses**

†National Renewable Energy Lab, Golden, CO 80401

**ACKNOWLEDGMENT**

The authors acknowledge center for computational science and technology services at Tulane University for CCS cluster access. We also want to thank Bocarsly group in Princeton University and Donahue Group in Tulane University for helpful discussions.